\documentclass[aps,prd,10pt,showpacs,amsmath,showkeys,twocolumn,floatfix,amssymb, preprintnumbers, nofootinbib, superscriptaddress]{revtex4-1} 
\usepackage{epsfig,dcolumn}
\usepackage{graphicx}
\usepackage{comment} 
\usepackage[usenames]{color}
\usepackage{graphicx}
\usepackage{bm}
 \usepackage{ifpdf}
\usepackage{subcaption}

\usepackage[normalem]{ulem}
\usepackage[dvipsnames]{xcolor}
\usepackage[utf8]{inputenc}
\usepackage{hyperref}
\hypersetup{ 
  pdfnewwindow=true,      
  colorlinks=true,        
  linkcolor=PineGreen,    
  citecolor=PineGreen,    
  filecolor=PineGreen,    
  urlcolor=PineGreen      
}

\newcommand{\be}{\begin{eqnarray}}
\newcommand{\ee}{\end{eqnarray}}

\begin{document}

\title{    Threshold expansion formula of $N$-boson  in finite volume from variational approach}

\author{Peng~Guo}
\email{pguo@csub.edu}

\affiliation{Department of Physics and Engineering,  California State University, Bakersfield, CA 93311, USA}
\affiliation{Kavli Institute for Theoretical Physics, University of California, Santa Barbara, CA 93106, USA}

\date{\today}

\begin{abstract} 
In present work, we show   how the threshold expansion formula of $N$ identical bosons in  finite volume may be derived by   iterations of Faddeev-type coupled dynamical equations.
The energy shift of $N$-boson system near threshold  is dominated by zero momenta mode of  $N$-body amplitudes with  all particles nearly static. The dominant zero momenta mode and sub-leading non-zero momenta mode contributions  are connected through finite volume Faddeev-type coupled dynamical equations. Eliminating non-zero momenta modes by iterations ultimately yields an analytic expression that can be solved by threshold expansion.
  \end{abstract}

\maketitle

\section{Introduction}
\label{intro} 
Quantum mechanical many-body dynamics is essential  for the understanding of  wide range   phenomena in modern physics, including Bose-Einstein condensate and superfluidity  \cite{London1938Phenomenon,Landau:1941vsj,Bogolyubov:1947zz}. The   many-body dynamics usually rely on approximate approaches in the past, such as Hartree-Fock method \cite{Hartree1935}. In recent years, a lot progresses  have been made toward the study of few- and many-body dynamics from first principle, quantum chromodynamics (QCD)  \cite{Aoki:2007rd,Feng:2010es,Lang:2011mn,Aoki:2011yj,Dudek:2012gj,Dudek:2012xn,Wilson:2014cna,Wilson:2015dqa,Dudek:2016cru,Beane:2007es,Horz:2019rrn,Detmold:2008fn,Detmold:2008yn}. The calculation of lattice QCD   is usually performed  in Euclidean space with all particles   confined in a periodic cubic box, hence   the multihadron dynamics  is not directly accessible.  Instead dynamics is encoded in a discrete energy spectrum of   multihadron system in finite volume. 
Therefore, establishing a method of  mapping out infinite-volume multihadron dynamics from discrete energy spectrum in finite volume has become an important subject in past few years.  Such a connection in two-body sector  is established   by L\"uscher formula in \cite{Luscher:1990ux} and its extensions \cite{Rummukainen:1995vs,Christ:2005gi,Bernard:2008ax,He:2005ey,Lage:2009zv,Doring:2011vk,Briceno:2012yi,Hansen:2012tf,Guo:2012hv,Guo:2013vsa}. Many promising developments along different approaches   have been made toward few- and many-body finite volume systems  recently \cite{Kreuzer:2008bi,Kreuzer:2009jp,Kreuzer:2012sr,Polejaeva:2012ut,Briceno:2012rv,Hansen:2014eka,Hansen:2015zga,Hansen:2016fzj,Hammer:2017uqm,Hammer:2017kms,Meissner:2014dea,Briceno:2017tce,Sharpe:2017jej,Mai:2017bge,Mai:2018djl, Doring:2018xxx, Romero-Lopez:2018rcb,Guo:2016fgl,Guo:2017ism,Guo:2017crd,Guo:2018ibd,Guo:2018xbv,Guo:2019hih,Blanton:2019igq,Romero-Lopez:2019qrt,Guo:2019ogp,Blanton:2019vdk,Mai:2019fba}.  One crucial thing to justify these  recent developments is to perform some tests and reproduce some known results, such as the threshold expansion formula that was originally derived by perturbation theory \cite{Huang:1957im,Lee:1957zzb,Beane:2007qr,Detmold:2008gh}.

Motivated exactly   by the purpose of  testing our formalism on finite volume $N$-body dynamics based on variational approach \cite{Guo:2018ibd,Guo:2019hih,Guo:2019ogp}, in this work, we illustrate how the well-known  threshold expansion formula for $N$-identical-boson system \cite{Huang:1957im,Lee:1957zzb,Beane:2007qr,Detmold:2008gh}  may be derived from  coupled dynamical equations. The exact value of eigen-energy of $N$-body system  are given by the eigen-solution of these Faddeev-type  coupled dynamical equations. Faddeev-type coupled dynamical equations is a non-perturbative approach, hence it applies in principle to both weakly and strongly coupled system. To reproduce  threshold expansion formula, the perturbation expansion in terms of weak coupling is carried out by iterations of coupled dynamical equations. A energy dependent  closed form    is thus obtained, and it ultimately yields the  threshold expansion formula by further expansion near threshold. The threshold expansion formula up to $\mathcal{O} (\eta^4/L^6)$ for pair-wise interaction and   $\mathcal{O} (\eta_3/L^6 )$ for three-body interaction  is already known \cite{Beane:2007qr,Detmold:2008gh}, where $\eta$ and $\eta_3$ are the two-body   and   three-body coupling strengths respectively.   The exact expression of $\mathcal{O} (\eta^4/L^6)$ expansion formula requires higher order terms by multiple iterations, which ultimately becomes a tedious task.  To simplify our presentation since the result is not new, in this work, we will only show  the derivation of the threshold expansion formula  up to  $\mathcal{O} (\eta^3/L^5)$ and  $\mathcal{O} (\eta_3/L^6 )$ by a single iteration, in terms of perturbation theory, they may be associated with   $\eta^2$ and $\eta \eta_3$ order diagrams respectively.

The paper is organized as follows.  The   formalism of finite volume  $N$-identical-boson systems is presented   in detail  in Section \ref{Nbdynamics}.  The derivation of threshold expansion formula  is illustrated       in Section \ref{ThresholdExpansion}.     Summary  is  given in Section \ref{summary}.

\section{ $N$-boson dynamics in finite volume }\label{Nbdynamics} 

The dynamics of $N$ non-relativistic identical bosons in finite volume is described by Lippmann-Schwinger type integral equation, see Refs.~\cite{Guo:2019hih,Guo:2019ogp},
\begin{equation} 
\Phi_E(\{\mathbf{ x}\})  = \int_{L^3} \prod_{i=1}^N d \mathbf{ x}'_i G_E (\{\mathbf{ x} - \mathbf{ x}'\} )  V(\{\mathbf{ x}'\})  \Phi_E(\{\mathbf{ x}'\}) , \label{LSeq}
\end{equation} 
where the position of i-th particle is denoted by $\mathbf{ x}_i$, and $\{\mathbf{ x}\}  = \{\mathbf{ x}_1, \cdots , \mathbf{ x}_N \}$. The $N$-body finite volume Green's function is given by
\begin{equation}
G_E (\{\mathbf{ x}  \} )    = \frac{1}{L^{3 N}} \sum_{ \{ \mathbf{ p}\}} \frac{ e^{ i \sum_{i=1}^N  \mathbf{ p}_i \cdot \mathbf{ x}_i } }{E - \sum_{i=1}^N  \frac{\mathbf{ p}^2_i}{2m} },
\end{equation} 
where $ \{ \mathbf{ p}\} =  \{ \mathbf{ p}_1, \cdots , \mathbf{ p}_N\}$, and $\mathbf{ p}_i = \frac{2\pi}{L} \mathbf{ n}_i$ with $\mathbf{ n}_i \in \mathbb{Z}^3$ stands for the free momentum of i-th particle. $L$ is the size of the cubic box. The finite volume Green's function is the solution of differential equation,
\begin{equation}
\left ( E + \sum_{i=1}^N  \frac{ \nabla^2_i}{2m} \right )G_E (\{\mathbf{ x}  \} ) =\prod_{i=1}^N \sum_{\mathbf{ n}_i \in \mathbb{Z}^3} \delta( \mathbf{ x}_i + \mathbf{ n}_i L),
\end{equation}
and satisfies periodic boundary condition,
\begin{equation}
G_E (\{\mathbf{ x}  + \mathbf{ n} L \} ) = G_E (\{\mathbf{ x}    \} ).
\end{equation}
Due to the periodic nature of Green's function, the periodicity of wave function, $\Phi_E(\{\mathbf{ x} + \mathbf{ n} L\}) =\Phi_E(\{\mathbf{ x}\}) $, is hence automatically warranted by Eq.(\ref{LSeq}). The interactions among particles is described by $V(\{\mathbf{ x}\})$,    the same form as given in  \cite{Beane:2007qr} is used in present work, {\it i.e.} only contact pair-wise and three-body interactions are considered,
\begin{equation}
V(\{\mathbf{ x}\}) = \eta \sum_{(i<j)=1}^N \delta (\mathbf{ r}_{i j}) +  \eta_3 \sum_{(i<j<k)=1}^N \delta (\mathbf{ r}_{i k})\delta (\mathbf{ r}_{j k}),
\end{equation}
where $ \mathbf{ r}_{i j} =\mathbf{ x}_i - \mathbf{ x}_j$ is relative coordinate between i-th and j-th particles, and $\eta$ and $\eta_3$ are the coupling strengths for pair-wise and three-body contact interactions respectively.

As illustrated in Refs.~\cite{Guo:2019hih,Guo:2019ogp}, two types of finite volume Faddeev amplitudes may be introduced by
\begin{equation}
T_{(i j)} ( \{ \mathbf{ k}\}) = -  \int_{L^3}  \left ( \prod_{l=1}^N  d \mathbf{ x}_l e^{-  i    \mathbf{ k}_l \cdot \mathbf{ x}_l } \right )  \eta \delta (\mathbf{ r}_{i j})\Phi_E(\{\mathbf{ x}\})  ,
\end{equation}
and
\begin{align}
& T_{(ij k )} ( \{ \mathbf{ k}\})  \nonumber \\
& = -  \int_{L^3}  \left ( \prod_{l=1}^N  d \mathbf{ x}_l e^{-  i    \mathbf{ k}_l \cdot \mathbf{ x}_l } \right )  \eta_3 \delta (\mathbf{ r}_{i k})\delta (\mathbf{ r}_{j k})\Phi_E(\{\mathbf{ x}\})  ,
\end{align}
where  $T_{(i j)}$'s and  $T_{(i j k )} $'s are associated with pair-wise and three-body contact interactions respectively. There are totally $\frac{N(N-1)}{2}$   $T_{(i j)}$'s and  $\frac{N(N-1) (N-2)}{6}$  $T_{(i j k )} $'s.  Eq.(\ref{LSeq}) is thus turned into $\frac{(N+1) N (N-1)}{6}$ coupled equations, for instance,
\begin{align}
 & T_{(12)}   ( \{ \mathbf{ k}\})  = \frac{1}{L^3} \sum_{   \mathbf{ p}_2  } \frac{ \eta  }{E - \frac{   \sum_{i=1}^N \mathbf{ p}^2_i}{2m}  } \nonumber \\
& \times \left [  \sum_{i<j} T_{(ij)} ( \{  \mathbf{ p} \} ) + \sum_{i<j<k} T_{(ij k)} ( \{ \mathbf{ p} \} )  \right ],   \nonumber \\
&    \mathbf{ p}_1=  \mathbf{ k}_1 +  \mathbf{ k}_2 -\mathbf{ p}_2 , \ \   \mathbf{ p}_l= \mathbf{ k}_l,  \ \  l=3, \cdots, N, 
\end{align}
and
\begin{align}
 & T_{(123)}  ( \{ \mathbf{ k}\})    = \frac{1}{L^6} \sum_{   \mathbf{ p}_2,\mathbf{ p}_3 } \frac{  \eta_3  }{E - \frac{   \sum_{i=1}^N\mathbf{ p}^2_i  }{2m}   }  \nonumber \\
 & \times \left [   \sum_{i<j} T_{(ij)} ( \{  \mathbf{ p} \} ) + \sum_{i<j<k} T_{(ij k)} ( \{ \mathbf{ p} \} ) \right ],   \nonumber \\
 &     \mathbf{ p}_1=   \sum_{i=1}^3 \mathbf{ k}_i  -\mathbf{ p}_2 - \mathbf{ p}_3 , \ \   \mathbf{ p}_l= \mathbf{ k}_l,  \ \  l=4, \cdots, N. 
\end{align} 
The rest of equations for $T_{(i j)}$'s and  $T_{(i j k )} $'s are thus obtained by swapping particle indices: $ 1 \leftrightarrow i$, $ 2 \leftrightarrow j$ and $ 3 \leftrightarrow k$.

\subsection{Symmetry consideration}
\begin{figure}
\begin{center}
\includegraphics[width=0.48\textwidth]{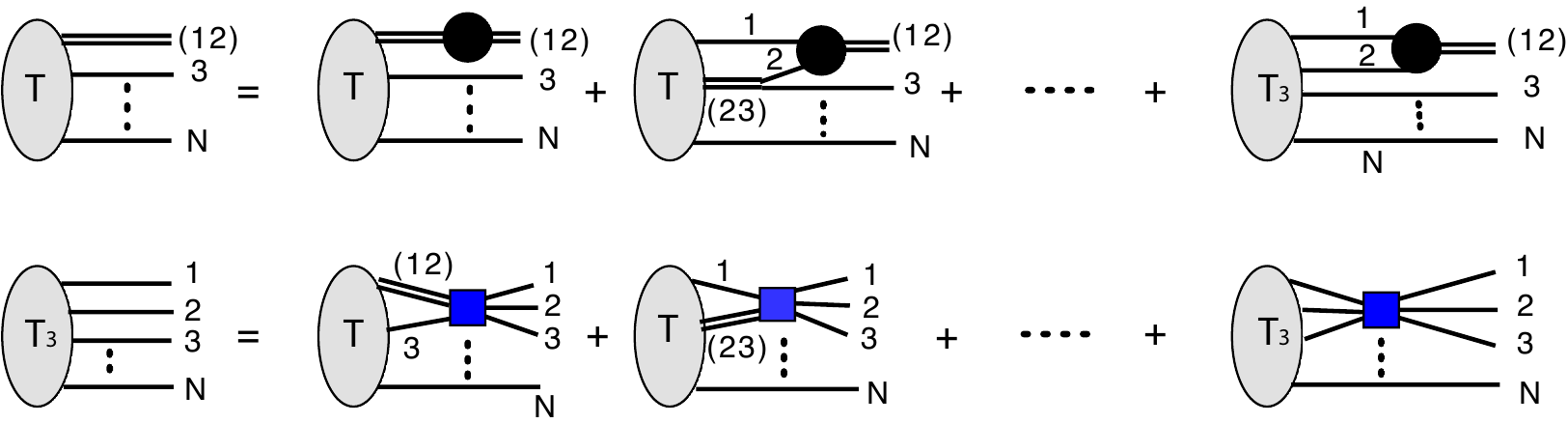}
\caption{  Diagrammatic representation of Eq.(\ref{2bLSeq}) and Eq.(\ref{3bLSeq}), pair-wise and three-body interactions are represented by black solid circle and blue solid square.   }\label{feynmanplot}
\end{center}
\end{figure}

Because of exchange symmetry of $N$ identical bosons system, only two independent amplitudes are required.  Let's define
\begin{equation}
T  ( \overline{\{ \mathbf{ k}\}}_{(12)}) = T_{(12)} ( \{ \mathbf{ k}\}) , \label{2bamp}
\end{equation}
where  $\overline{\{ \mathbf{ k}\}}_{(12)} = \{    \mathbf{ k}_3, \cdots, \mathbf{ k}_N  \}$ is a subset of $ \{ \mathbf{ k}\} = \{    \mathbf{ k}_1,  \mathbf{ k}_2,  \overline{\{ \mathbf{ k}\}}_{(12)}  \}$ by removing first two elements, and 
\begin{equation}
 T_{ 3} (\overline{ \{ \mathbf{ k}\}}_{(123)} )  =  T_{(123 )} ( \{ \mathbf{ k}\}), \label{3bamp}
\end{equation}
where  $\overline{\{ \mathbf{ k}\}}_{(123)} = \{   \mathbf{ k}_4,  \cdots, \mathbf{ k}_N  \}$ is a subset of $ \{ \mathbf{ k}\} = \{    \mathbf{ k}_1,  \mathbf{ k}_2,   \mathbf{ k}_3,  \overline{\{ \mathbf{ k}\}}_{(123)}  \}$ by removing first three elements.   According to 
    Eq.(\ref{2bamp}), $T_{(12)} ( \{ \mathbf{ k}\}) $    in fact depends on both $\mathbf{ k}_1 + \mathbf{ k}_2$ and $\overline{\{ \mathbf{ k}\}}_{(12)} $,    the $\mathbf{ k}_1 + \mathbf{ k}_2$ dependence has been dropped due to the fact that   all momenta are constrained by momentum conservation $\sum_{i=1}^N  \mathbf{ k}_i = \mathbf{ P}$, where $\mathbf{ P}$ stands for total momentum of N-particle.  Similarly,   the $\mathbf{ k}_1 + \mathbf{ k}_2 +  \mathbf{ k}_3 $ dependence in $ T_{(123 )} ( \{ \mathbf{ k}\})$ is dropped as well because of momentum conservation constraint. The rest of amplitudes are related to $T$ and $T_3$ defined in Eqs.(\ref{2bamp}) and (\ref{3bamp}) respectively by
\begin{equation}
T_{(i j)} ( \{ \mathbf{ k}\})  = T  ( \overline{\{ \mathbf{ k}\}}_{(ij)}), \ \   \ \ \ \  T_{(ij k )} ( \{ \mathbf{ k}\}) =   T_3  ( \overline{\{ \mathbf{ k}\}}_{(ij k)}),
\end{equation}
where $ \overline{\{ \mathbf{ k}\}}_{(ij)}$ and $ \overline{\{ \mathbf{ k}\}}_{(ij k)}$ can be obtained from 
sets $ \overline{\{ \mathbf{ k}\}}_{(12)}$ and $ \overline{\{ \mathbf{ k}\}}_{(123)}$ by swapping particle momenta: $ \mathbf{ k}_1 \leftrightarrow \mathbf{ k}_i$, $ \mathbf{ k}_2 \leftrightarrow \mathbf{ k}_j$ and $ \mathbf{ k}_3 \leftrightarrow \mathbf{ k}_k$.  Two sets of coupled equations for $T_{(i j)}$'s and  $T_{(i j k )} $'s are hence reduced to two equations,
\begin{align}
 & T  ( \overline{\{ \mathbf{ k}\}}_{(12)})  \nonumber \\
 & = \frac{\eta}{L^3} \sum_{   \mathbf{ p}_2  } \frac{ \sum_{i<j} T ( \overline{\{  \mathbf{ p} \}}_{(ij)}) + \sum_{i<j<k} T_{3} ( \overline{\{ \mathbf{ p} \}}_{(ij k)})   }{E - \frac{  ( \mathbf{ k}_1 +  \mathbf{ k}_2 - \mathbf{ p}_2 )^2  +   \mathbf{ p}_2^2 + \sum_{i=3}^N\mathbf{ k}^2_i }{2m}   }  ,   \nonumber \\
 &    \mathbf{ p}_1=  \mathbf{ k}_1 +  \mathbf{ k}_2 -\mathbf{ p}_2 , \ \   \mathbf{ p}_l= \mathbf{ k}_l,  \ \  l=3, \cdots, N,   \label{2bLSeq}
\end{align}
  and
\begin{align}
 & T_{3} (\overline{ \{ \mathbf{ k}\}}_{(123)} )   \nonumber \\
 &= \frac{\eta_3}{L^6} \sum_{   \mathbf{ p}_2,\mathbf{ p}_3 }  \frac{   \sum_{i<j} T  ( \overline{\{  \mathbf{ p} \}}_{(ij)}) + \sum_{i<j<k} T_{3} ( \overline{\{ \mathbf{ p} \}}_{(ijk)})    }{E - \frac{( \sum_{i=1}^3 \mathbf{ k}_i -   \mathbf{ p}_2- \mathbf{ p}_3  )^2 +    \mathbf{ p}_2^2 + \mathbf{ p}_3^2 +  \sum_{i=4}^N\mathbf{ k}^2_i  }{2m}   }  ,    \nonumber \\
 &     \mathbf{ p}_1=   \sum_{i=1}^3 \mathbf{ k}_i  -\mathbf{ p}_2 - \mathbf{ p}_3 , \ \   \mathbf{ p}_l= \mathbf{ k}_l,  \ \  l=4, \cdots, N,  \label{3bLSeq}
\end{align}
where   $\{ \mathbf{ p}\} = \{ \mathbf{ p}_1 ,\mathbf{ p}_2,  \overline{\{ \mathbf{ p}\}}_{(12)}\}  =  \{ \mathbf{ p}_1 ,\mathbf{ p}_2, \mathbf{ p}_3,  \overline{\{ \mathbf{ p}\}}_{(123)}\} $ in both Eqs.(\ref{2bLSeq}) and (\ref{3bLSeq}). The diagrammatic representation of  Eq.(\ref{2bLSeq}) and Eq.(\ref{3bLSeq}) is given  in Fig.~\ref{feynmanplot}.

\subsection{Three-boson dynamical equations}
In the case of $N=3$, the dynamical equations   are  thus given by
\begin{equation}
  T  (   \mathbf{ k}_3 )   = \frac{\eta}{L^3} \sum_{   \mathbf{ p}_2  } \frac{  T (  \mathbf{ k}_3 ) + 2 T (  \mathbf{ p}_2 )    + T_{3}  }{E - \frac{ ( \mathbf{ k}_1 +  \mathbf{ k}_2 - \mathbf{ p}_2 )^2  +  \mathbf{ p}_2^2 +  \mathbf{ k}^2_3 }{2m}   }  ,  
\end{equation}
and
\begin{equation}
  T_{3}   = \frac{\eta_3}{L^6} \sum_{   \mathbf{ p}_2,\mathbf{ p}_3 }  \frac{   3  T (  \mathbf{ p}_2 )   +  T_{3}     }{E - \frac{  \mathbf{ p}_2^2 + \mathbf{ p}_3^2 + (   \mathbf{ P} -   \mathbf{ p}_2- \mathbf{ p}_3  )^2   }{2m}   }  .  
\end{equation}
Eliminating $T_3$ amplitude, we find
\begin{align}
&  T  (   \mathbf{ k}_3 )  
  = \frac{\eta}{L^3} \sum_{   \mathbf{ p}_2  } \frac{  1     }{E - \frac{  \mathbf{ p}_2^2 + ( \mathbf{ k}_1 +  \mathbf{ k}_2 - \mathbf{ p}_2 )^2  +  \mathbf{ k}^2_3 }{2m}   }   \Bigg [  T  (   \mathbf{ k}_3 )  + 2 T (  \mathbf{ p}_2 )  \nonumber \\
&  \quad \quad \quad   + \frac{ \frac{\eta_3}{L^6} \sum_{   \mathbf{ q}_2,\mathbf{ q}_3 }  \frac{   3  T (  \mathbf{ q}_2 )      }{E - \frac{  \mathbf{ q}_2^2 + \mathbf{ q}_3^2 + (  \mathbf{ P} -   \mathbf{ q}_2- \mathbf{ q}_3  )^2   }{2m}   }  }{1- \frac{\eta_3}{L^6} \sum_{   \mathbf{ q}_2,\mathbf{ q}_3 }  \frac{   1    }{E - \frac{  \mathbf{ q}_2^2 + \mathbf{ q}_3^2 + (  \mathbf{ P} -   \mathbf{ q}_2- \mathbf{ q}_3  )^2   }{2m}   }} \Bigg ].   \label{N3Teq}
\end{align}

\section{Threshold Expansion}\label{ThresholdExpansion}
In this section, we illustrate that the threshold expansion formula may be derived from Eqs.(\ref{2bLSeq}) and (\ref{3bLSeq}) by iterations. Near the ground state energy threshold, all $N$ particles  are nearly at rest for weak interactions. Hence the dominant contribution comes from the zero momenta mode of amplitudes: $\{ \mathbf{ k}\} = \{ \mathbf{ 0}\}$. Thus, we find
\begin{align}
 & \left [ 1-  \frac{\eta}{L^3} \sum_{   \mathbf{ p}_2  } \frac{1 }{E - \frac{  \mathbf{ p}_2^2     }{m}   }  \right ] T  ( \overline{\{ \mathbf{ 0}\}}_{(12)})  \nonumber \\
 & = \frac{\eta}{L^3} \sum_{   \mathbf{ p}_2  } \frac{ \sum_{i<j}^{ (ij) \neq (12) } T ( \overline{\{  \mathbf{ p} \}}_{(ij)}) + \sum_{i<j<k} T_{3} ( \overline{\{ \mathbf{ p} \}}_{(ijk)})   }{E - \frac{  \mathbf{ p}_2^2     }{m}   }  ,      \nonumber \\
 &   \quad  \mathbf{ p}_1=   -\mathbf{ p}_2 , \ \   \mathbf{ p}_l= \mathbf{ 0},  \ \  l=3, \cdots, N,  \label{2bzeromode}
\end{align}
and
\begin{align}
 & T_{3} (\overline{ \{ \mathbf{ 0}\}}_{(123)} )   \nonumber \\
 &= \frac{\eta_3}{L^6} \sum_{   \mathbf{ p}_2,\mathbf{ p}_3 }  \frac{   \sum_{i<j} T  ( \overline{\{  \mathbf{ p} \}}_{(ij)}) + \sum_{i<j<k} T_{3} ( \overline{\{ \mathbf{ p} \}}_{(ijk)})    }{E - \frac{  \mathbf{ p}_2^2 + \mathbf{ p}_3^2 + (   \mathbf{ p}_2+ \mathbf{ p}_3  )^2   }{2m}   }  ,    \nonumber \\
 &   \quad   \mathbf{ p}_1=     -\mathbf{ p}_2 - \mathbf{ p}_3 , \ \   \mathbf{ p}_l= \mathbf{ 0},  \ \  l=4, \cdots, N  .  
\end{align}

\subsection{Perturbation expansion by iteration of $N$-body dynamical equations }\label{iteration}
As we can see from above equations, the leading order contributions of $T$ and $T_3$  start at the order of  $ \frac{\eta}{L^3} $  and $ \frac{\eta_3}{L^6} $ respectively. In present work, the aim is to just simply illustrate how the threshold expansion formula are derived from Eqs.(\ref{2bLSeq}) and (\ref{3bLSeq}).  For this purpose, we will only compute up to $  \eta^3/L^5 $ order in threshold expansion formula by iterating Eqs.(\ref{2bLSeq}) and (\ref{3bLSeq}) only once. The contributions from three-body force are only kept at the lowest order effect,  and also splitting  up zero momenta mode and non-zero momenta mode  contributions  in Eq.(\ref{2bzeromode}), so we obtain
\begin{align}
 & \left [ 1-  \frac{\eta}{L^3} \sum_{   \mathbf{ p}_2 \neq \mathbf{ 0}  } \frac{1 }{E - \frac{  \mathbf{ p}_2^2     }{m}   }  -  \frac{\eta}{L^3}  \frac{ \frac{N(N-1)}{2} }{E} \right ] T  ( \overline{\{ \mathbf{ 0}\}}_{(12)}) \nonumber \\
 & =  \frac{\eta}{L^3} \sum_{   \mathbf{ p}_2 \neq \mathbf{ 0}  } \frac{ \sum_{i<j}^{ (ij) \neq (12) } T ( \overline{\{  \mathbf{ p} \}}_{(ij)})   }{E - \frac{  \mathbf{ p}_2^2     }{m}   }  \nonumber \\
 &  +    \frac{\eta}{L^3} \frac{\frac{N (N-1)(N-2)}{6}}{E}  T_{3} ( \overline{\{ \mathbf{ 0} \}}_{(123)})    + \mathcal{O} (\frac{\eta^2 \eta_3}{L^{12}})  ,  \nonumber \\
 & \quad  \mathbf{ p}_1 = - \mathbf{ p}_2 , \ \ \ \ \mathbf{ p}_l =\mathbf{ 0}, \ \ l=3, \cdots, N,        \label{nonzeromode}
\end{align}
and
\begin{align}
 & T_{3} (\overline{ \{ \mathbf{ 0}\}}_{(123)} )  = \frac{\eta_3}{L^6}   \frac{     \frac{N(N-1)}{2}    }{E    } T  ( \overline{\{ \mathbf{ 0}\}}_{(12)}) + \mathcal{O} (\frac{\eta \eta_3}{L^{9}}) .  
\end{align}
Eliminating $T_{3} (\overline{ \{ \mathbf{ 0}\}}_{(123)} ) $ term in Eq.(\ref{nonzeromode}), we find
\begin{align}
 & \left [ 1-  \frac{\eta}{L^3} \sum_{   \mathbf{ p}_2 \neq \mathbf{ 0}  } \frac{1 }{E - \frac{  \mathbf{ p}_2^2     }{m}   }  -  \frac{\eta}{L^3}  \frac{ \frac{N(N-1)}{2} }{E}     \right ] T  ( \overline{\{ \mathbf{ 0}\}}_{(12)}) \nonumber \\
 &  =     \frac{\eta \eta_3 }{L^9} \frac{\frac{N^2 (N-1)^2(N-2)}{12}}{E^2}     T  ( \overline{\{ \mathbf{ 0}\}}_{(12)})   \nonumber \\
 &+   \frac{\eta}{L^3} \sum_{   \mathbf{ p}_2 \neq \mathbf{ 0}  } \frac{ \sum_{i<j}^{ (ij) \neq (12) } T ( \overline{\{  \mathbf{ p} \}}_{(ij)})   }{E - \frac{  \mathbf{ p}_2^2     }{m}   }  + \mathcal{O} (\frac{\eta^2 \eta_3}{L^{12}})  ,   \nonumber \\
 & \quad  \mathbf{ p}_1 = - \mathbf{ p}_2 , \ \ \ \ \mathbf{ p}_l =\mathbf{ 0}, \ \ l=3, \cdots, N.        \label{T2nonzeromode}
\end{align}
Now, dominant zero momenta mode and sub-leading non-zero momenta mode are well separated in Eq.(\ref{T2nonzeromode}).
The terms that are given by non-zero momenta mode of $T$ amplitudes in Eq.(\ref{T2nonzeromode})  can be eliminated   and thus are related to zero momenta mode amplitudes by iterating Eq.(\ref{2bLSeq})  once.

Non-zero momenta mode of set $ \overline{\{  \mathbf{ p} \}}_{(ij)}$ in $T ( \overline{\{  \mathbf{ p} \}}_{(ij)})  $ can be split into two groups: (1)  $ \overline{\{  \mathbf{ p} \}}_{(1j )} = \{ \mathbf{ 0}, \cdots , \mathbf{ p}_2 , \cdots , \mathbf{ 0} \}$ with only a single non-zero momentum dependence at j-th position, $ \mathbf{ p}_j = \mathbf{ p}_2$  and  $j >2 $; (2) $ \overline{\{  \mathbf{ p} \}}_{(i j )} = \{ \mathbf{ 0}, \cdots, - \mathbf{ p}_2 ,  \cdots , \mathbf{ p}_2 , \cdots , \mathbf{ 0} \}$ with   two non-zero momenta  dependence at i-th and j-th positions,  $ \mathbf{ p}_i =- \mathbf{ p}_2$ and $ \mathbf{ p}_j = \mathbf{ p}_2$, where $ (i<j) = 3, \cdots, N$.

(1) For amplitudes in group one with only a single non-zero momentum dependence, using Eq.(\ref{2bLSeq}) again, we find  that each non-zero mode $ T ( \overline{\{  \mathbf{ p} \}}_{(1j)})$ is related to two   amplitudes that doesn't depend on $\mathbf{ p}_2$, 
\begin{align}
& T ( \overline{\{  \mathbf{ p} \}}_{(1j)})   \nonumber \\
& =  \frac{\eta}{L^3}  \left [  \sum_{   \mathbf{ q}_2  } \frac{  T ( \overline{\{  \mathbf{ q} \}}_{(1j)})       }{E - \frac{   \mathbf{ q}_1^2  +   \mathbf{ q}_2^2 +  \mathbf{ p}^2_2 }{2m}   }  +  \sum_{   \mathbf{ q}_1  } \frac{    T ( \overline{\{  \mathbf{ q} \}}_{(2 j)})    }{E - \frac{   \mathbf{ q}_1^2  +   \mathbf{ q}_2^2 +  \mathbf{ p}^2_2 }{2m}   }   \right ] + \cdots \nonumber \\
& =  \frac{\eta}{L^3} \sum_{   \mathbf{ q}_2  } \frac{  2 T ( \overline{\{  \mathbf{ q} \}}_{(1j)})      }{E - \frac{  ( \mathbf{ p}_2 + \mathbf{ q}_2 )^2  +   \mathbf{ q}_2^2 +  \mathbf{ p}^2_2 }{2m}   }  + \cdots , \label{groupone}   
\end{align}
where $ \overline{\{  \mathbf{ q} \}}_{(1j)} = \{ \mathbf{ 0}, \cdots , \mathbf{ q}_2, \cdots, \mathbf{ 0} \}$   and  $ \overline{\{  \mathbf{ q} \}}_{(2j)} = \{ \mathbf{ 0}, \cdots ,   \mathbf{ q}_1, \cdots, \mathbf{ 0} \}$   with $\mathbf{ q}_2$ and  $ \mathbf{ q}_1$ siting at j-th position.  Splitting sum of $\mathbf{ q}_2$ to zero momenta mode and non-zero momenta mode again in Eq.(\ref{groupone}), the dominant contribution for $ T ( \overline{\{  \mathbf{ p} \}}_{(1j)})$  comes from   zero momenta mode, sub-leading contribution from non-zero momenta mode may be eliminated by iteration again. Keeping only dominant zero momenta mode contribution,  we hence find
\begin{equation}
 T ( \overline{\{  \mathbf{ p} \}}_{(1j)})  = \frac{\eta}{L^3} \frac{1}{E - \frac{\mathbf{ p}_2^2}{m}} 2  T ( \overline{\{  \mathbf{ 0} \}}_{(12)})  +  \mathcal{O} (\frac{\eta \eta_3}{L^{6}}).
\end{equation}
There are $ (N-2)$  terms of $ T ( \overline{\{  \mathbf{ p} \}}_{(1j)})$  in group one, and another $(N-2)$  equivalent terms for     $ T ( \overline{\{  \mathbf{ p} \}}_{(2j)})$ amplitudes.  Therefore, the total dominant contribution from group one is  $ \frac{\eta}{L^3} \frac{4 (N-2)}{E - \frac{\mathbf{ p}_2^2}{m}}   T ( \overline{\{  \mathbf{ 0} \}}_{(12)})$.

(2) For amplitudes in group two with two non-zero momentum dependence, each  $ T ( \overline{\{  \mathbf{ p} \}}_{(i j)})$     is related to only one   amplitude that does not depend on $ \mathbf{ p}_i =- \mathbf{ p}_2$ and $ \mathbf{ p}_j = \mathbf{ p}_2$,
\begin{equation}
 T ( \overline{\{  \mathbf{ p} \}}_{(i j)})  
 =  \frac{\eta}{L^3}    \sum_{   \mathbf{ q}_2  } \frac{  T ( \overline{\{  \mathbf{ q} \}}_{(i j)})       }{E - \frac{    \mathbf{ q}_2^2 +  \mathbf{ p}^2_2 }{m}   } + \cdots   , 
\end{equation}
where $ \overline{\{  \mathbf{ q} \}}_{(i j)} = \{ \mathbf{ 0}, \cdots ,  - \mathbf{ q}_2, \cdots,   \mathbf{ q}_2, \cdots, \mathbf{ 0} \}$     with $ - \mathbf{ q}_2$ and  $ \mathbf{ q}_2$ siting at i-th and j-th positions respectively.  Hence, the dominant zero momenta mode contribution from  $ T ( \overline{\{  \mathbf{ p} \}}_{(i j)})$ term is 
\begin{equation}
 T ( \overline{\{  \mathbf{ p} \}}_{(i j)})    = \frac{\eta}{L^3} \frac{1}{E - \frac{\mathbf{ p}_2^2}{m}}   T ( \overline{\{  \mathbf{ 0} \}}_{(12)})  +  \mathcal{O} (\frac{\eta \eta_3}{L^{6}}). 
\end{equation}
There are  $\frac{(N-2)(N-3)}{2}$ such terms,   the total   number of dominant contribution from group two is thus  $ \frac{\eta}{L^3} \frac{\frac{(N-2)(N-3)}{2}}{E - \frac{\mathbf{ p}_2^2}{m}}   T ( \overline{\{  \mathbf{ 0} \}}_{(12)})$.

\subsubsection{Zero momenta mode $N$-boson dynamical equation}
Combining all non-zero momenta mode terms in Eq.(\ref{T2nonzeromode}) from both group one and group two, we obtain,
\begin{align}
 &    \frac{\eta}{L^3} \sum_{   \mathbf{ p}_2 \neq \mathbf{ 0}  } \frac{ \sum_{i<j}^{ (ij) \neq (12) } T ( \overline{\{  \mathbf{ p} \}}_{(ij)})   }{E - \frac{  \mathbf{ p}_2^2     }{m}   }  \nonumber \\
 &  =  \frac{\eta^2}{L^6} \sum_{   \mathbf{ p}_2 \neq \mathbf{ 0}  } \frac{ 4 (N-2) + \frac{(N-2) (N-3)}{2}  }{ \left ( E - \frac{  \mathbf{ p}_2^2     }{m}  \right )^2  }    + \mathcal{O} (\frac{\eta^2 \eta_3}{L^{12}})  .  
\end{align}
Plugging  them back into Eq.(\ref{nonzeromode}),   we thus find 
\begin{align}
 & \left [ 1-  \frac{\eta}{L^3} \sum_{   \mathbf{ p}_2 \neq \mathbf{ 0}  } \frac{1 }{E - \frac{  \mathbf{ p}_2^2     }{m}   }  -  \frac{\eta}{L^3}  \frac{ \frac{N(N-1)}{2} }{E} \right ] T  ( \overline{\{ \mathbf{ 0}\}}_{(12)}) \nonumber \\
 &  =     \frac{\eta \eta_3}{L^9}  \frac{\frac{N^2 (N-1)^2(N-2)}{12}}{E^2}   T  ( \overline{\{ \mathbf{ 0}\}}_{(12)})     \nonumber \\
 &  + \frac{\eta^2}{L^6} \sum_{   \mathbf{ p}_2 \neq \mathbf{ 0}  } \frac{   \frac{(N-2) (N+5)}{2}  }{ \left ( E - \frac{  \mathbf{ p}_2^2     }{m}  \right )^2  }      T  ( \overline{\{ \mathbf{ 0}\}}_{(12)})   + \mathcal{O} (\frac{\eta^2 \eta_3}{L^{12}})  .   \label{LO}
\end{align}
 Zero momenta mode amplitude $   T  ( \overline{\{ \mathbf{ 0}\}}_{(12)})  $  is thus cancelled out from both sides of equation, and Eq.(\ref{LO}) yields an analytic   form that depends on only $E$ and momentum sum.

\subsubsection{Three-boson example}
Using three-body dynamical equation given in Eq.(\ref{N3Teq}) as a specific example, setting $\mathbf{ k}_1 = \mathbf{ k}_2 = \mathbf{ k}_3 =\mathbf{ 0}$, and keep only up to $ \frac{\eta \eta_3}{L^9}$ order, we obtain
\begin{equation}
  T  (   \mathbf{ 0} )  
  \simeq \frac{\eta}{L^3} \sum_{   \mathbf{ p}_2  } \frac{   T  (   \mathbf{ 0} )  + 2 T (  \mathbf{ p}_2 )     }{E - \frac{  \mathbf{ p}_2^2  }{m}   }        + \frac{\eta \eta_3}{L^9}   \frac{   3  T (  \mathbf{ 0} )      }{E^2   }  .  \label{ThreebodyT2example} 
\end{equation}
Splitting up to zero momenta and non-zero momenta mode in  Eq.(\ref{ThreebodyT2example}), and also use Eq.(\ref{N3Teq}) once to eliminate non-zero momenta mode,  
\begin{equation}
  T  (   \mathbf{ p}_2 )  
  = \frac{\eta}{L^3}  \frac{   2 T (  \mathbf{ 0} )      }{E - \frac{    \mathbf{ p}^2_2 }{m}   }    + \cdots,  \ \ \mathbf{ p}_2 \neq \mathbf{ 0},
\end{equation}
 hence we finally get 
\begin{align}  
& 1  - \frac{\eta}{L^3} \sum_{   \mathbf{ p}_2  \neq \mathbf{ 0} } \frac{   1  }{E - \frac{  \mathbf{ p}_2^2  }{m}   } -  \frac{\eta}{L^3} \frac{3}{E}   \nonumber \\
& \simeq   \frac{\eta^2}{L^6} \sum_{   \mathbf{ p}_2 \neq \mathbf{ 0}  } \frac{   4   }{ \left ( E - \frac{  \mathbf{ p}_2^2  }{m}  \right )^2  }    + \frac{\eta \eta_3}{L^9}   \frac{   3      }{E^2   }  .  
\end{align}

\subsection{Near threshold   expansion and ground state energy}\label{thresexpand}
By assuming that energy shift near threshold is small due to weak interactions:  $E\sim 0$, Eq.(\ref{LO}) is thus turned into a polynomial equation by   near threshold expansion, keeping up to $\mathcal{O} (E^3)$, we   have
\begin{align}
 & \mathcal{O} (E^4) + \left (  \frac{1}{L^3}  \sum_{   \mathbf{ p} \neq \mathbf{ 0}  } \frac{1}{ \frac{  \mathbf{ p}^4     }{m^2}   } \right )  E^3  \nonumber \\
 & +  \left ( \frac{1}{\eta}+  \frac{1}{L^3} \sum_{   \mathbf{ p} \neq \mathbf{ 0}  } \frac{1}{ \frac{  \mathbf{ p}^2     }{m}   } -  \frac{\eta}{L^6} \sum_{   \mathbf{ p} \neq \mathbf{ 0}  } \frac{\frac{(N+5) (N-2)}{2}}{ \frac{  \mathbf{ p}^4     }{m^2}   }   \right ) E^2  \nonumber \\
 &  -  \frac{1}{L^3}    \frac{N(N-1)}{2}  E   =     \frac{ \eta_3}{L^9}  \frac{N^2 (N-1)^2(N-2)}{12}      .    \label{threshold}
\end{align}
Introducing renormalized two-body coupling constant 
\begin{equation}
\frac{1}{\eta} = \frac{1}{\eta_R} - \frac{ m \Lambda}{\pi L},
\end{equation}
where $\Lambda$ is related to the cutoff on momentum sum,  and also using  relations given in Refs. \cite{Beane:2007qr},
\begin{align}
& \frac{1}{L^3} \sum_{   \mathbf{ p}_1 \neq \mathbf{ 0}  } \frac{1}{ \frac{  \mathbf{ p}_1^2     }{m}   } - \frac{ m \Lambda}{\pi L} =  \frac{ m \mathcal{I}}{(2\pi)^2 L} , \ \ \ \ \mathcal{I} =   \sum_{ \mathbf{ n} \neq \mathbf{ 0}  }^{| \mathbf{ n} |  \leqslant \Lambda} \frac{1}{   \mathbf{ n}^2     } -  4\pi  \Lambda , \nonumber \\
&  \frac{1}{L^3} \sum_{   \mathbf{ p}_1 \neq \mathbf{ 0}  } \frac{1}{ \frac{  \mathbf{ p}_1^4     }{m^2}   }  =  \frac{ m^2 L \mathcal{J}}{(2\pi)^4 } , \ \  \ \ \mathcal{J} =   \sum_{ \mathbf{ n} \neq \mathbf{ 0}  }  \frac{1}{   \mathbf{ n}^4     } ,
\end{align}
we can rewrite Eq.(\ref{threshold}) to
\begin{align}
 &  \frac{   L \mathcal{J}}{(2\pi)^4 }    (mE)^3  \nonumber \\
 & +  \left ( \frac{1}{ m \eta_R}+   \frac{  \mathcal{I}}{(2\pi)^2 L}  - \frac{(N+5) (N-2)}{2}   \frac{  m \eta_R   \mathcal{J}}{(2\pi)^4 L^2 }   \right ) (m E)^2  \nonumber \\
 &  -  \frac{1}{L^3}    \frac{N(N-1)}{2}  (m E)   \simeq    \frac{ m \eta_3}{L^9}  \frac{N^2 (N-1)^2(N-2)}{12}      .    \label{cubiceq}
\end{align}
The cubic equation, Eq.(\ref{cubiceq}), can be easily solved by perturbation theory
\begin{equation}
m E= \frac{N(N-1)}{2} \frac{4\pi a_0}{L^3}  \left [ 1 +\sum_{n=1}^3  \left (\frac{a_0}{ \pi L}  \right )^n c_n   \right ],
\end{equation}
where $a_0$ is two-body scattering length and is related to coupling constant of pair-wise contact  interaction  by  
\begin{equation}
 m \eta_R = 4\pi a_0 .
 \end{equation}
    The solution of cubic equation, Eq.(\ref{cubiceq}) is thus given by
\begin{align}
 E &= \frac{N(N-1)}{2} \frac{ 4\pi a_0 }{ m L^3} \bigg [ 1 - \left  ( \frac{   a_0 }{ \pi L}  \right  ) \mathcal{I}   \nonumber \\
 &+  \left ( \frac{ a_0 }{\pi L} \right   )^2 \left ( \mathcal{I}^2 +(2 N-5) \mathcal{J}  \right ) +  \mathcal{O} (  \frac{   a^3_0 }{L^3}      )  \bigg]  \nonumber \\
 & + \frac{N(N-1) (N-2)}{6} \frac{ \eta_3}{L^6}  ,
\end{align}
 which is consistent with well-known results in Refs.~\cite{Huang:1957im,Lee:1957zzb,Beane:2007qr,Detmold:2008gh}.

\begin{figure*}
\centering
\begin{subfigure}[b]{0.32\textwidth}
\includegraphics[width=\textwidth]{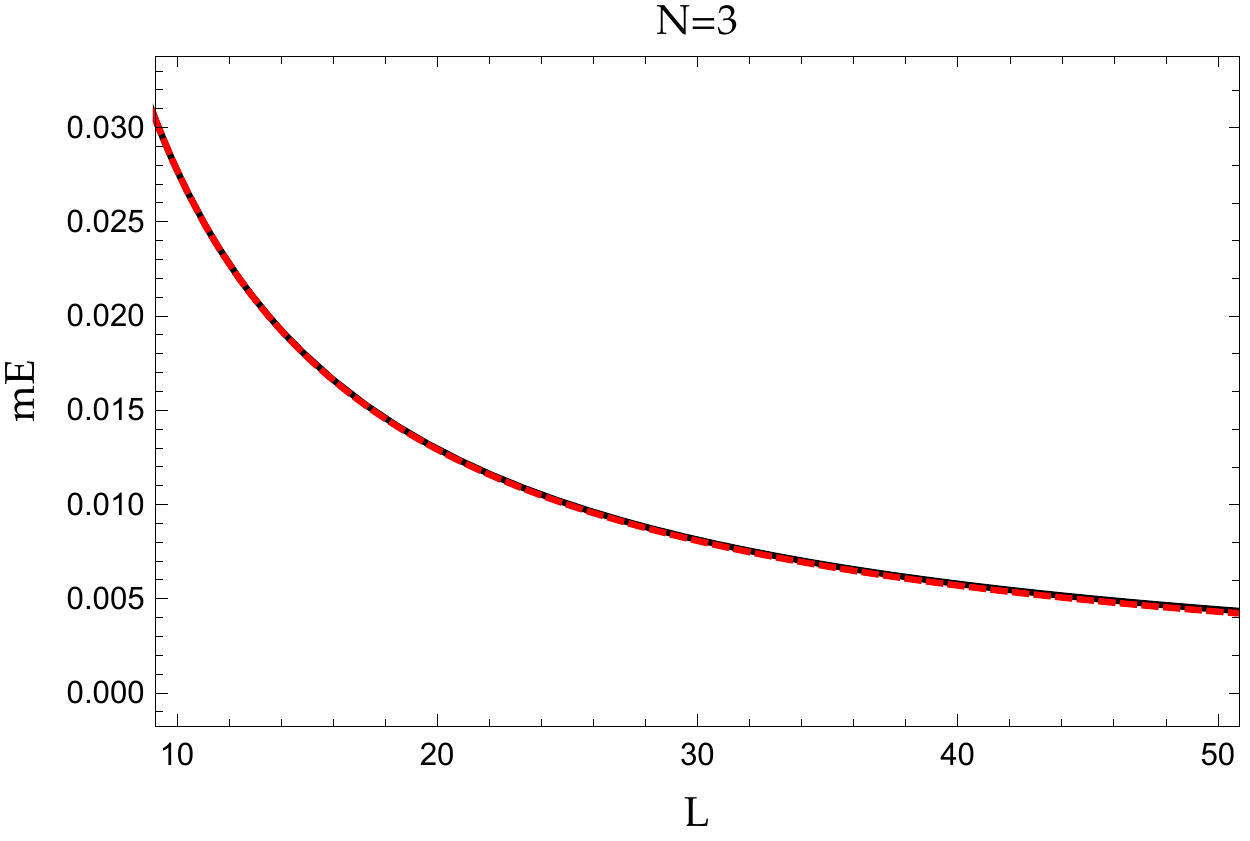}
\caption{ Three-boson energy spectrum   in 1D.}
\end{subfigure}
\begin{subfigure}[b]{0.32\textwidth}
\includegraphics[width=\textwidth]{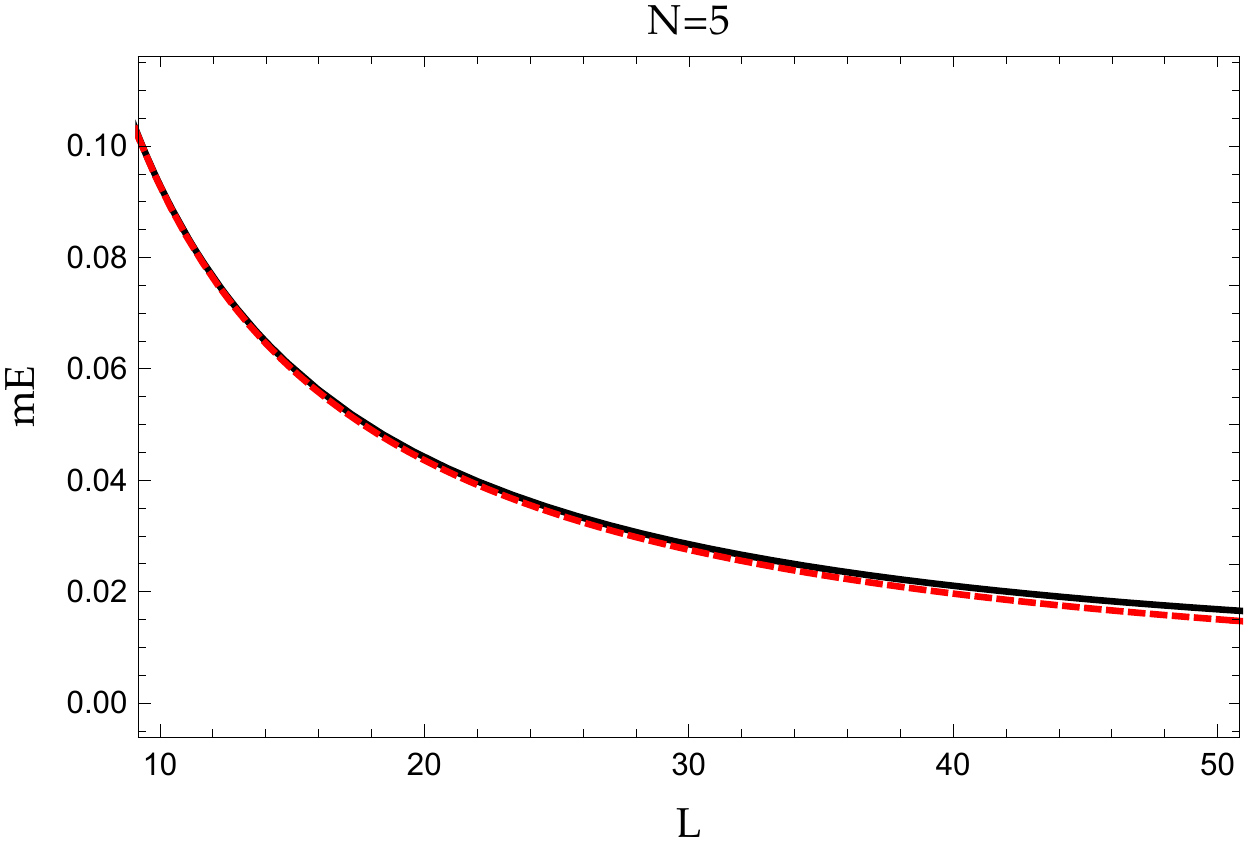}
\caption{ Five-boson energy spectrum   in 1D.}
\end{subfigure}
\begin{subfigure}[b]{0.32\textwidth}
\includegraphics[width=\textwidth]{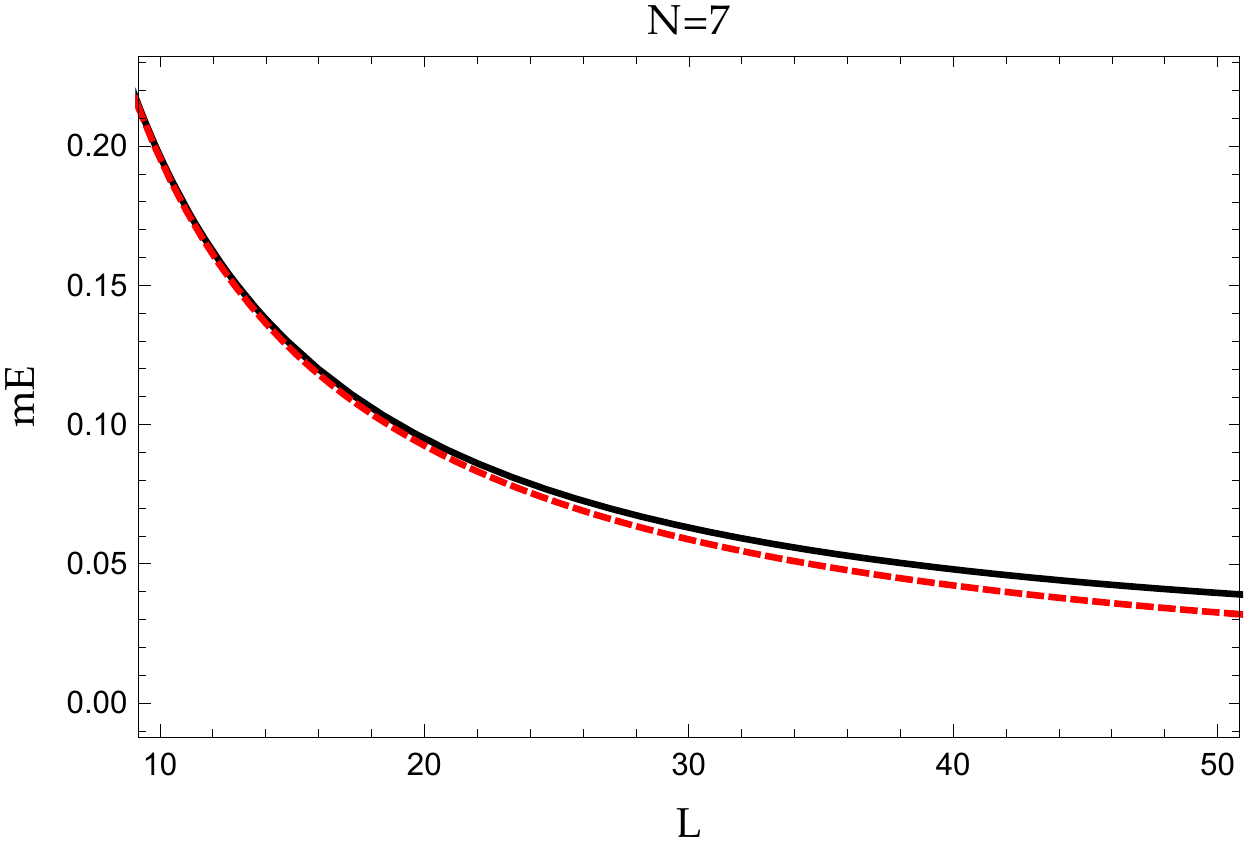}
\caption{ Seven-boson energy spectrum   in 1D.}
\end{subfigure}
\caption{ Plot of  $m E$ as the function of $L$: red dashed curves are  exact solutions given by Eq.(\ref{exact1D}), and black solid curves are approximate solution given by threshold expansion in Eq.(\ref{approxE1D}),  $m \eta = 0.1$.  }
\label{ENbplot}
\end{figure*}

\subsection{Threshold expansion formula in 1D and comparison to exact solutions} 
 The iteration of coupled finite volume $N$-body dynamical equation approach  and results presented in section \ref{iteration} and  \ref{thresexpand} can be applied to $N$-boson interaction in 1D with little   changes. $N$ identical bosons interacting with pair-wise contact potentials in 1D is in fact exactly solvable, see Refs.~\cite{Yang:1967bm,Lieb:1963rt,Guo:2016fgl}. The exact analytic solutions are given by $E = \frac{1}{2m} \sum_{i=1}^N p_i^2$, where $p_i$'s satisfies  coupled equations 
\begin{equation}
\frac{p_i L}{2} = \sum_{j =1 (j \neq i)}^N \cot^{-1}  \left ( \frac{p_i - p_j}{m \eta} \right ), \ \ \sum_{i=1}^N p_i=0. \label{exact1D}
\end{equation}

 Keeping only pair-wise contact interaction and expanding up to $\mathcal{O} (E^3)$, the  threshold   expansion equation in 1D can be obtained by replacing 3D momentum sum $\frac{1}{L^3} \sum_{\mathbf{ p}} $ in Eq.(\ref{threshold}) by 1D counterpart $\frac{1}{L} \sum_{p} $, hence, we obtain
  \begin{align}
 & \frac{1}{L}    \frac{N(N-1)}{2}     \simeq \left (  \frac{1}{L}  \sum_{  p \neq  0  } \frac{1}{  p^4        } \right )  (mE)^2  \nonumber \\
 & +  \left ( \frac{1}{ m \eta}+  \frac{1}{L } \sum_{  p \neq  0 } \frac{1}{  p^2      } -  \frac{\eta}{L^2} \sum_{   p \neq  0 } \frac{\frac{(N+5) (N-2)}{2}}{   p^4       }   \right ) (mE)   .    \label{threshold1D}
\end{align}
The infinite momentum sum in 1D can be carried out rather easily,  
\begin{equation}
 \sum_{  p \neq  0  } \frac{1}{  p^2        }= \frac{L^2}{12}, \ \ \ \ \sum_{  p \neq  0  } \frac{1}{  p^4        }= \frac{L^4}{720},
\end{equation}
the solution of Eq.(\ref{threshold1D}) is thus given by
\begin{equation}
m E= \frac{ \sqrt{ b^2  + 4 c  }   -  b}{2}  , \label{approxE1D}
\end{equation}
where 
\begin{align}
b &= \frac{720}{L^3}  \left ( \frac{1}{ m \eta}+  \frac{L}{12} -   \frac{(N+5) (N-2)}{2}  \frac{m \eta  L^2}{720}  \right )  , \nonumber \\
  c &= \frac{N(N-1)}{2}  \frac{720}{L^4} .
\end{align}
The comparison of $m E$ as the function of $L$ between exact solutions given by Eq.(\ref{exact1D}) and approximate solution by threshold expansion in Eq.(\ref{approxE1D}) is illustrated in  Fig.~\ref{ENbplot}.

\section{Summary }\label{summary}

As a sanity check and a test on the formalism of finite volume $N$-body system developed in  \cite{Guo:2018ibd,Guo:2019hih,Guo:2019ogp}, we illustrate   how the well-known threshold expansion formula of   $N$-identical-boson system   may be derived by   iterations of Faddeev-type coupled dynamical equations. The ground state energy of $N$-boson system   near threshold  is dominated by zero momenta mode of  $N$-body amplitudes, non-zero momenta mode amplitudes are associated with sub-leading order contributions and are related to leading order zero momenta mode     through  Faddeev-type coupled dynamical equations. Eliminating non-zero momenta modes by iterations ultimately yields an   analytic expression that   depends on only system energy and free momentum sum,  thus it can be turned into a polynomial equation by treating energy shift near threshold as a small parameter.  With only a single iteration, we are  able to compute  threshold expansion formula up to  $\mathcal{O} (\eta^3/L^5)$ for pair-wise interaction and   $\mathcal{O} (\eta_3/L^6 )$ for three-body interaction.

\bigskip

\begin{acknowledgements}
 We   acknowledge support from the Department of Physics and Engineering, California State University, Bakersfield, CA.   This research was supported in part by the National Science Foundation under Grant No. NSF PHY-1748958.   We also thank M.~D\"oring for suggesting such an investigation. 
  \end{acknowledgements}

\appendix

\bibliography{ALL-REF.bib}

\end{document}